# Pathological Analysis of Blood Cells Using Deep Learning Techniques

*Virender Ranga[a], Shivam Gupta[b], Priyansh Agrawal[a], Jyoti Meena[a]

[a]*Department of Computer Engineering, National Institute of Technology , Kurukshetra , Haryana, India*
[b]*Department of Computer Science and Engineering, Indian Institute of Information Technology , Sonepat ,Haryana , India*

**Abstract—** Pathology deals with the practice of discovering the reasons for disease by analyzing the body samples. The most used way in this field, is to use histology which is basically studying and viewing microscopic structures of cell and tissues. The slide viewing method is widely being used and converted into digital form to produce high resolution images. This enabled the area of deep learning and machine learning to deep dive into this field of medical sciences. In the present study, a neural based network has been proposed for classification of blood cells images into various categories. When input image is passed through the proposed architecture and all the hyper parameters and dropout ratio values are used in accordance with proposed algorithm, then model classifies the blood images with an accuracy of 95.24%. The performance of proposed model is better than existing standard architectures and work done by various researchers. Thus model will enable development of pathological system which will reduce human errors and daily load on laboratory men. This will in turn help pathologists in carrying out their work more efficiently and effectively.

## 1. INTRODUCTION

The major area of work of pathologists is concerned with detecting the diseases and helping the patients in their healthcare and well-being. The present method used by pathologists for this purpose is manually viewing the slides using a microscope and other instruments [1]. But this method suffers from a lot of problems like there is no standard way of diagnosing, human errors and it puts a heavy load on the laboratory men to diagnose such a large number of slides daily[2]. In the nutshell this manual process is a very time taking, tiring and monotonous work. In the past few decades there has been a lot of advancement in the field of Artificial Intelligence. One of this is the growth of neural network namely convolutional neural network (CNN) [8]. Basically it uses the same copies of neuron which permits the network to have many neurons and help predicting large datasets and training models [7]. CNN have led to advance development in many pattern recognition, habit detection [3], video event summarization [4], facial expression recognition [5] and video surveillance [6]. Many standard CNN architectures listed in Table 1 have been proposed to solve standard classification problems. In the present study neural network is used to develop & train a model that will help in classification of blood cells images into various categories. These categories have significance in medical sciences. Thus this will help pathologists in carrying out their work efficiently and effectively.

## 2. LITERATURE REVIEW

Rosyadi et al. [18] carried out study that classified white cells images using Otsu's method for segmentation and k-mean-clusters for classification and were able to give 67% accurate results.

Gautam et al. also used naïve bayes to classify white blood cells using features like area, perimeter with an accuracy of 80.88% [19]. In chase for better performance Yu et al. devised a method using CNN for classification of blood cells. This method utilized various architectures like ResNet, InceptionV3, VGG16 & VGG19 [20]. The method performed at an accuracy of 88.5%. Filip Novoselnik in his paper used modified LeNet-5 consisting of 7 layers and achieved a performance of 81.11% [21]. Ahasan et al. using segmentation gave an accuracy of 88.57% [22]. Snehal Laddha used pre-trained convolutional network with multiclass models for SVM as feature extractor and classified white blood cells with an accuracy of 93.94% [23].

Table 1. STANDARD CNN ARCHITECTURES TIMELINE

| Architecture | Year Proposed |
|---|---|
| LeNet [9] | 1998 |
| AlexNet [10] | 2012 |
| VGG16 [11] | 2014 |
| VGG19 [11] | 2014 |
| ResNet-152 [12] | 2016 |
| Inception V3 [13] | 2016 |
| DenseNet-121 [14] | 2017 |
| DenseNet-169 [14] | 2017 |
| DenseNet-201 [14] | 2017 |
| MobileNet V1 [15] | 2017 |
| MobileNet V2 [16] | 2018 |
| NASNet [17] | 2019 |

*Corresponding email: virender.ranga@nitkkr.ac.in

### 2.1 *DATASET*

Database includes images of blood samples of healthy wellbeing [24]. The database had been provided to help in comparative evaluation of various techniques for classification of white blood cells. The dataset contains 12500 images of blood cell along with the labels .The cell types are polynuclear cells and mononuclear cells as shown in Fig. 1. and Fig. 2.

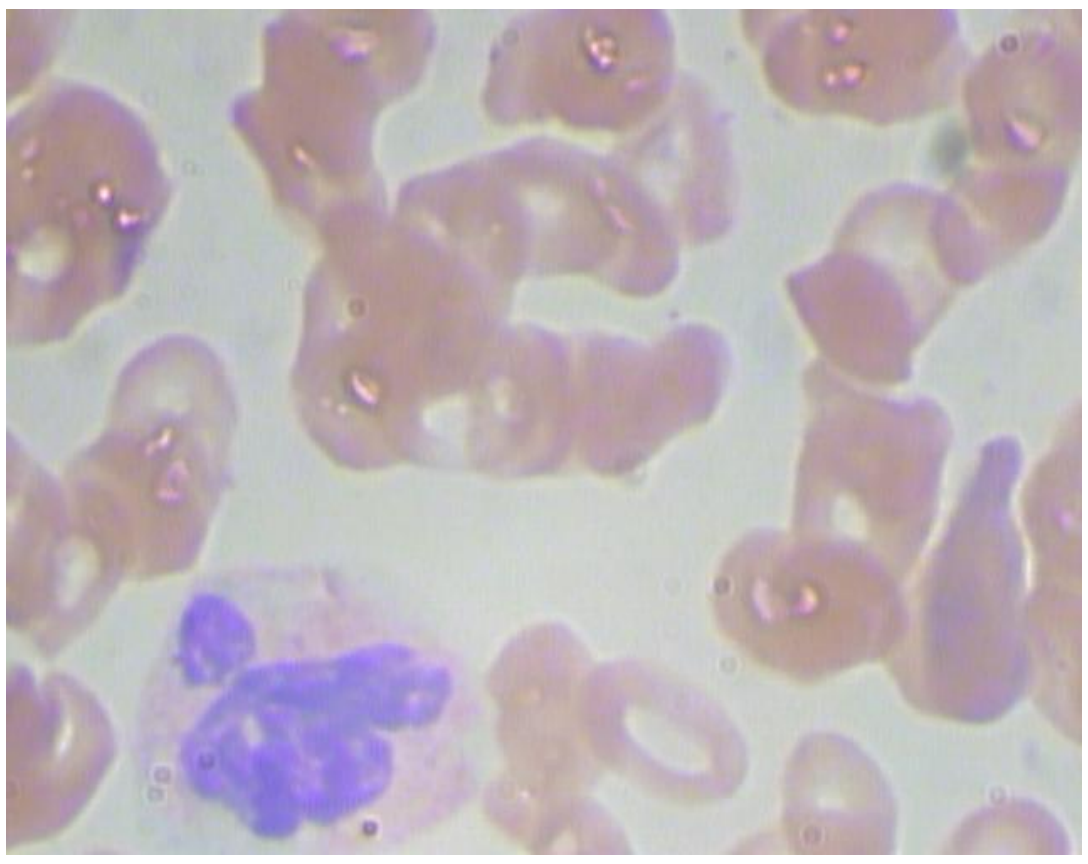

Fig.1. Mononuclear Cell

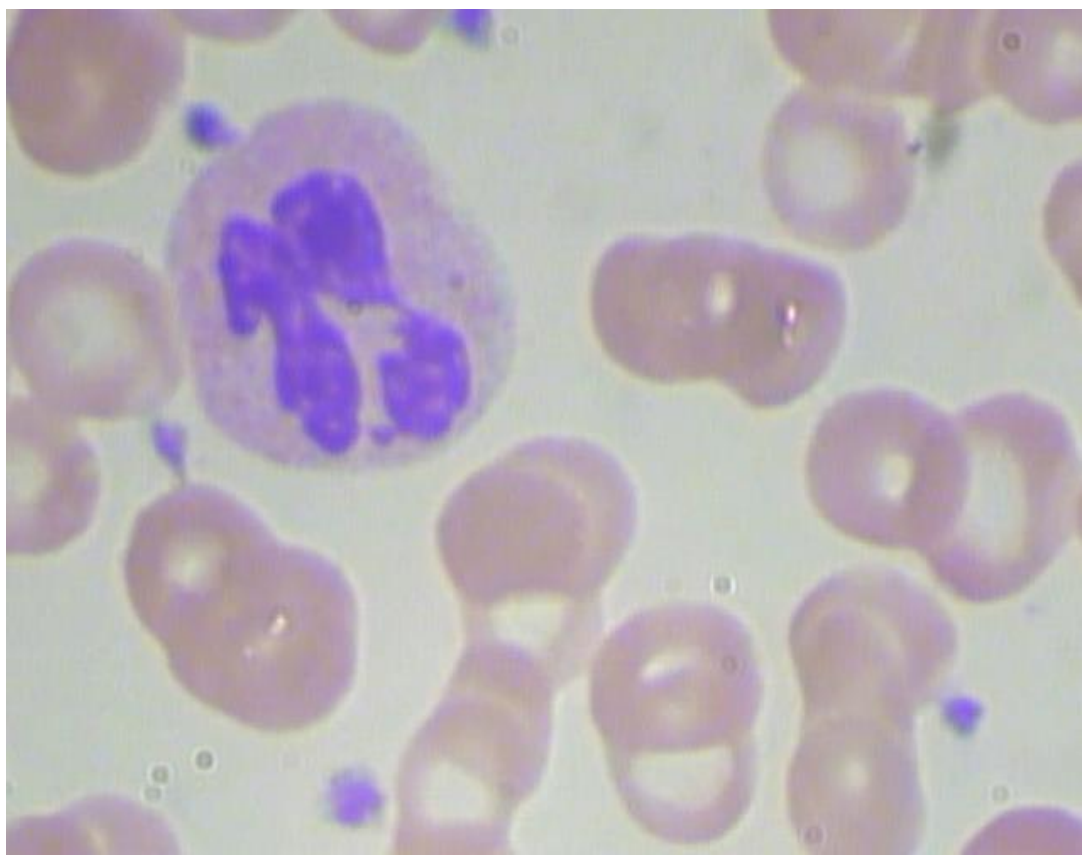

Fig. 2. Polynuclear Cell

## 3. RESEARCH METHODOLOGY

In the present study a neural network has been proposed. The proposed neural network is a sequential model with alternating three dimensional (3D) layers of convolution, pooling layer and an activation layer. The output from the 3D layers is modified to a 1D vector. The 1D vector is then converted into a dense fully connected neural network. After fully connected network a sigmoid function layer is used to get the probabilities of various classes.

### *3.1 Proposed Model Architecture*

The architecture for proposed neural network model is shown in Fig. 3. The architecture shows the sequence of various convolution, pooling and activation layers. Also the architecture shows the point of use of dropout layers.

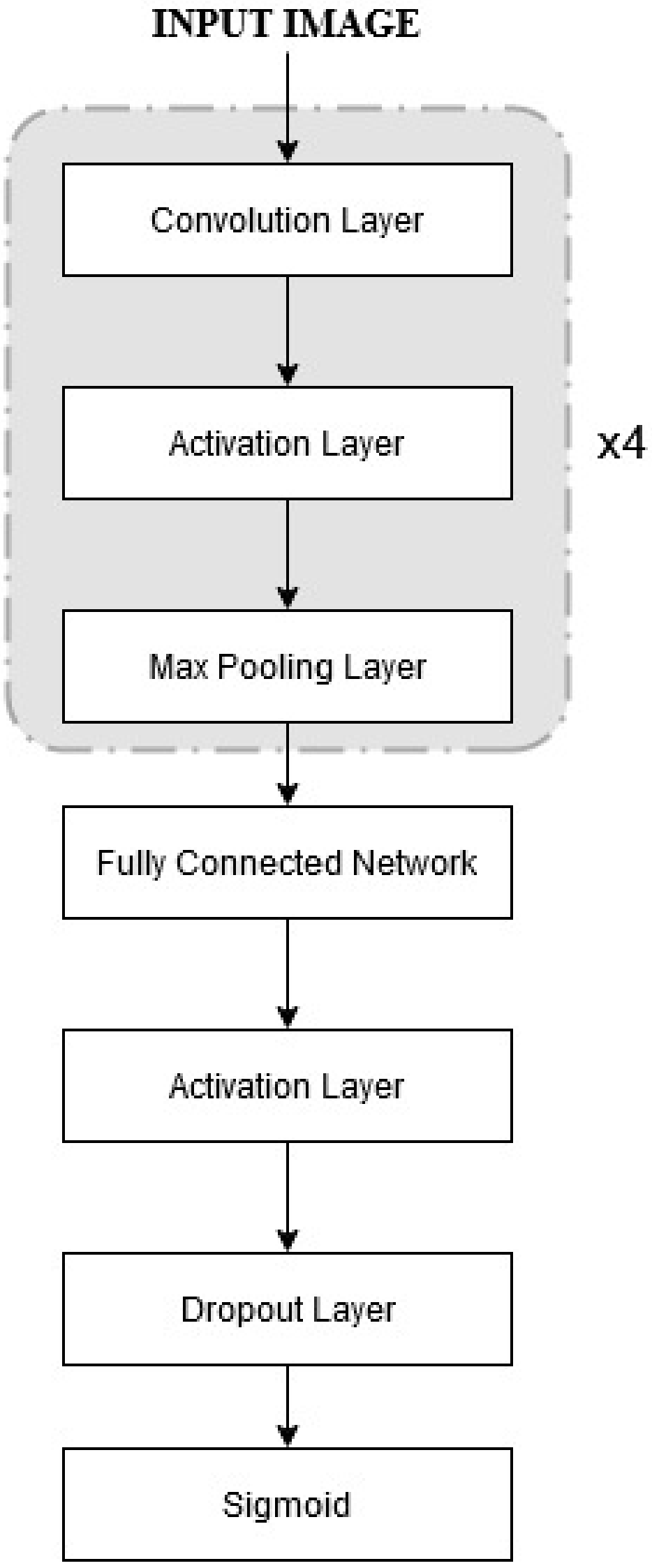

Fig. 3. Architecture of Proposed Model

### *3.2 Proposed Algorithm*

The pseudo code is presented below for the proposed neural network algorithm. The algorithm contains the details of hyper parameter values and dropout ratio values. These values will be used by of different layers in proposed model.

Step 1: Initialize a Sequential model S which take in input image I as 3D matrix.
Step 2: Add Convolution layer of filter size 3x3 with 32 neurons to S.
Step 3: Add Activation layer with ReLU function to S.
Step 4: Add Max Pooling layer with pool size of 2x2 to S.
Step 5: Add Convolution layer of filter size 3x3 with 32 neurons to S.
Step 6: Add Activation layer with ReLU function to S.
Step 7: Add Max Pooling layer with pool size of 2x2 to S.
Step 8: Add Convolution layer of filter size 3x3 with 32 neurons to S.
Step 9: Add Activation layer with ReLU function to S.
Step 10: Add Max Pooling layer with pool size of 2x2 to S.

Step 11: Add Convolution layer of filter size 3x3 with 64 neurons to S.
Step 12: Add Activation layer with ReLU function to S.
Step 13: Add Max Pooling layer with pool size of 2x2 to S.
Step 14: Convert three dimensional (3D) model S into a fully connected one dimensional (1D) model P.
Step 15: Reduce the number of neurons in P to 64 and maintain fully connected dense network.
Step 16: Add Activation layer with ReLU function to P.
Step 17: Apply a dropout ratio of 0.5 to neurons in P.
Step 18: Reduce the number of neurons in P to 2.
Step 19: Apply Activation function sigmoid to find out probabilities of each categories to be classified.
Step 20: Compile the model using loss function binary cross entropy with 'rmsprop' as optimizer. Use accuracy as metric.

### *3.3 Implementation of Proposed Model*

The proposed model is implemented using Keras library in python language. The model takes input image of size 120x160x3. The input is passed through a lambda layer that reduces the value of RGB matrix to a value between 0 and 1. After this the model is passed through layers as shown in the model architecture. The layers at each step use the hyper parameters and loss metrics as presented by proposed algorithm. The model is trained for 20 epochs on "*PARAM Shavak - Supercomputing Solution in a box*" with 4996 images of polynuclear cells and 4961 images of mononuclear cells .The trained model is saved in Hierarchical Data Format (h5 format) for future testing and practical use. The brief view of output shape after each layer and number of total parameters in each layer obtained during training of proposed model is shown in Table 2.

Table 2. BRIEF SUMMARY OF LAYERS IN PROPOSED MODEL

| Layer (type) | Output Shape | Param # |
|---|---|---|
| lambda_1 (Lambda) | (None, 120, 160, 3) | 0 |
| conv2d_1 (Conv2D) | (None, 118, 158, 32) | 896 |
| activation_1 (Activation) | (None, 118, 158, 32) | 0 |
| max_pooling2d_1 (MaxPooling2 | (None, 59, 79, 32) | 0 |
| conv2d_2 (Conv2D) | (None, 57, 77, 32) | 9248 |
| activation_2 (Activation) | (None, 57, 77, 32) | 0 |
| max_pooling2d_2 (MaxPooling2 | (None, 28, 38, 32) | 0 |
| conv2d_3 (Conv2D) | (None, 26, 36, 32) | 9248 |
| activation_3 (Activation) | (None, 26, 36, 32) | 0 |
| max_pooling2d_3 (MaxPooling2 | (None, 13, 18, 32) | 0 |
| conv2d_4 (Conv2D) | (None, 11, 16, 64) | 18496 |
| activation_4 (Activation) | (None, 11, 16, 64) | 0 |
| max_pooling2d_4 (MaxPooling2 | (None, 5, 8, 64) | 0 |
| flatten_1 (Flatten) | (None, 2560) | 0 |
| dense_1 (Dense) | (None, 64) | 163904 |
| activation_5 (Activation) | (None, 64) | 0 |
| dropout_1 (Dropout) | (None, 64) | 0 |
| dense_2 (Dense) | (None, 2) | 130 |
| activation_6 (Activation) | (None, 2) | 0 |

## 4. RESULTS AND DISCUSSIONS

After training the models on 20 epochs. The plots of training accuracy, testing accuracy and corresponding training loss, testing loss for proposed model is plotted using matplotlib and the trend so observed is provided below.

### *4.1 Proposed Model*

During training the accuracy of the proposed model follows a steady increase upto 20 epochs. The testing accuracy for the model at the beginning shows an up and down trend but it keeps on increasing steadily. Finally from 11th epochs the testing accuracy and training accuracy touches almost close value showing a good learning rate and making proposed model best suited for the problem. On the other hand model loss shows a reverse trend as it is complementary to accuracy. The final plot trend of model accuracy and loss is shown in Fig.4. and Fig. 5.

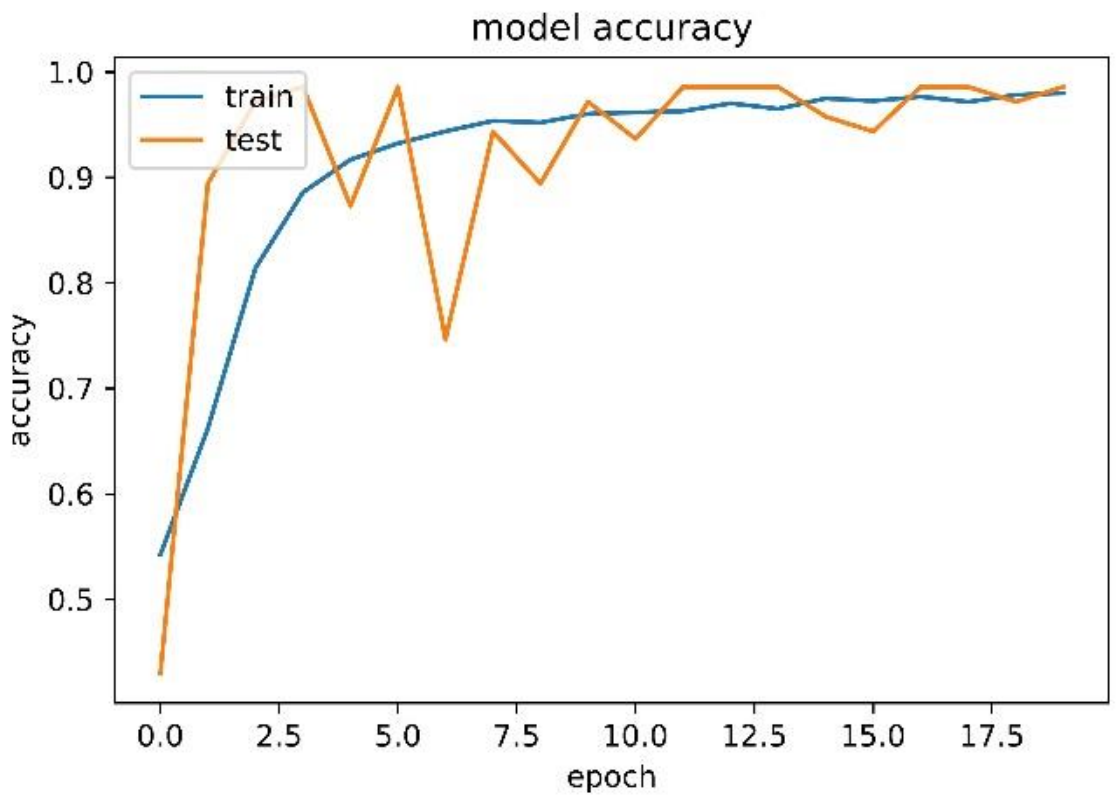

Fig. 4. Value of accuracy per epoch

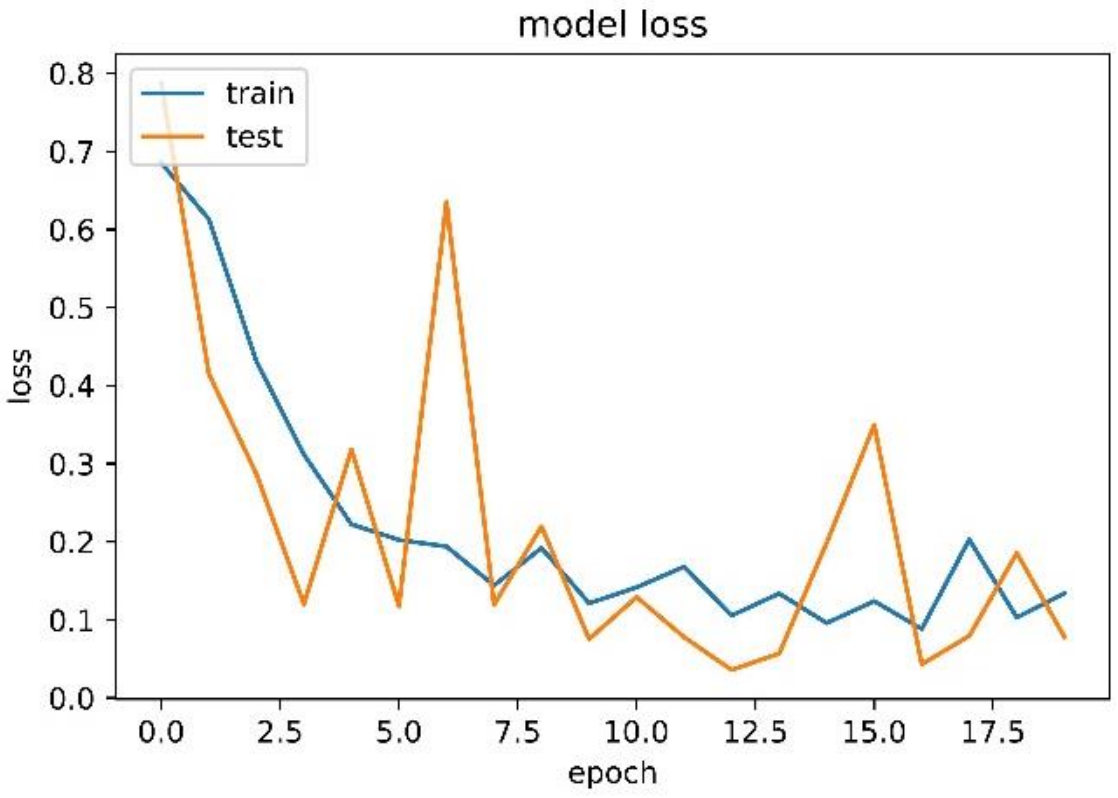

Fig. 5. Value of loss per epoch

### *4.1 Comparative Analysis with Standard Architectures*

The comparative analysis of standard convolutional neural network architectures with proposed model is shown in graphical form in Fig.6. The Fig.6 (a). compares the standard architectures on the basis of accuracy as metric. Accuracy of model determines quality of being correct or precise. The model having more accuracy is more precise. The two accuracies have been plotted in Fig.6 (a). namely training accuracy and testing accuracy. Training accuracy signifies how precise model is during training phase. Testing accuracy on other hand signifies how precise model performs when fed with unlearned input data.

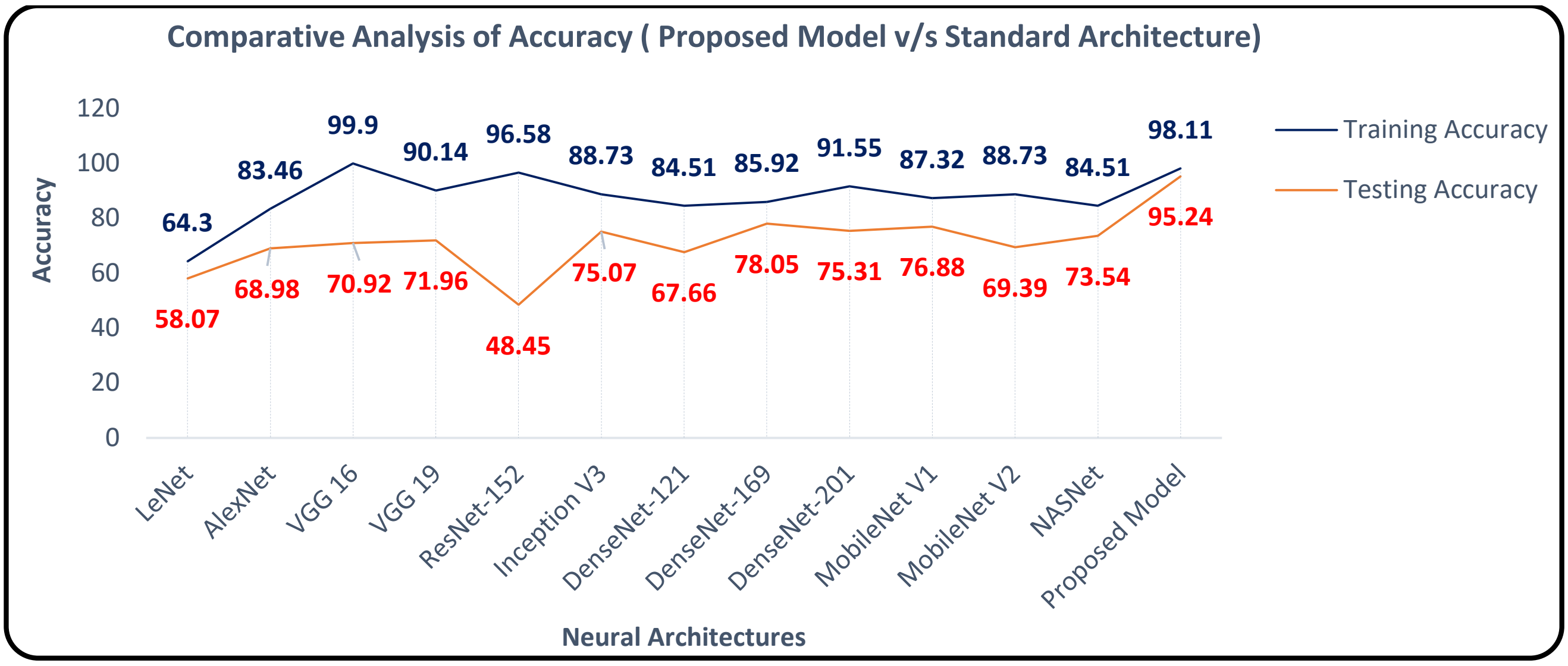

(a)

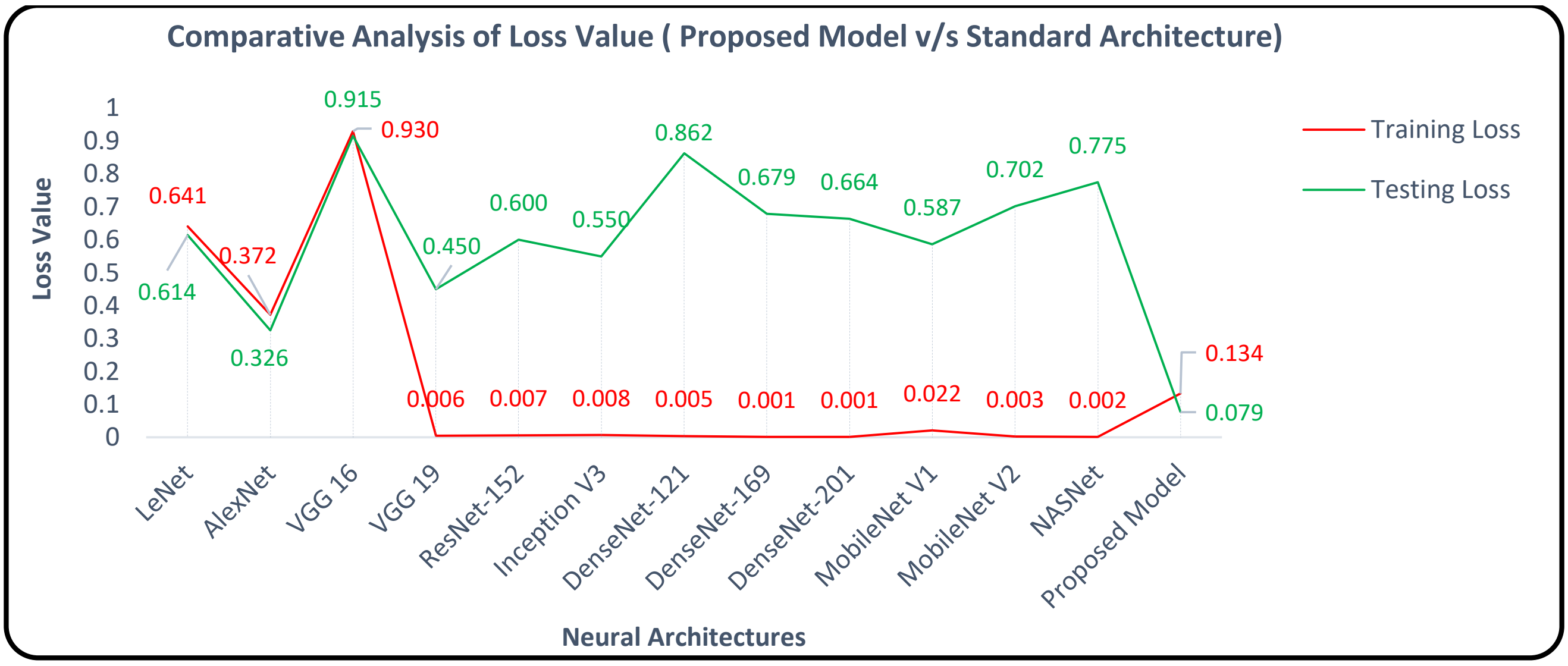

(b)

Fig. 6. Comparative analysis of proposed model with standard architectures using (a) Accuracy (b) Model Loss

From the graph it is clear that the proposed model is best fit model for our present study. Other standard architectures like VGG16, ResNet-152 may have training accuracy higher or close to proposed model but during testing the model, it doesn't perform upto the mark.

The comparison using loss value as metric, is shown in graphical form in Fig. 6 (b). Loss value of a model signifies how poorly a model behaves. Lower the loss better the model behaves. From Fig.6 (b). it is clear that various standard architectures may have lower training loss but when fed with new input image they don't perform well enough. The main reason for such a behavior is that they are not able to learn very well from the training data. This failure of learning can be due to their architecture and hyper parameters proposed in their algorithms. Thus in comparison to current standard CNN architecture, proposed model is performing better.

### *4.2 Comparative Analysis with Related Work*

The comparative analysis of proposed model with present related works carried out by various researchers is shown in the tabular form in Table 3. Thus it is clear that proposed model is having a testing accuracy of 95.24% which is better than all other present works.

Table 3. COMPARISON WITH RELATED WORK

| Method | Testing Accuracy |
|---|---|
| Rosyadi et al. | 67.00% |
| Gautam et al. | 80.88% |
| Filip et al. | 81.11% |
| Yu et al. | 88.50% |
| Ahsan et al. | 88.57% |
| Snehal et al. | 93.94% |
| **Proposed Model** | **95.24%** |

## 5. Conclusion

In the present study, a neural based network has been proposed for classification of blood cells images into various categories. These categories have significance in medical sciences. When input image is passed through the proposed architecture and all the hyper parameters and dropout ratio values are used in accordance with proposed algorithm, then model classifies the images with an accuracy of 95.24%. This accuracy is better than standard architectures listed in Table. 1. Further it can be seen that the proposed neural network performs better than present related works carried by various researchers. This can be clearly observed from Table. 3. Thus the proposed neural network architecture and algorithm will help in reducing human errors and daily load on laboratory men. This will also make the process of manual viewing less time taking and tiring. Therefore will help pathologists in carrying out their work efficiently and effectively.

## 6. Current and future development

Currently developed model classifies the cells into broadly two major groups of medical importance namely mononuclear and polynuclear. Models accuracy can be further improved using ensembling techniques of various CNN and RNN architectures. Even problem can be further sub-classified into neutrophil, monocyte, lymphocyte and eosinophil upon availability of large dataset for each of these classes.